\documentclass[baaa]{baaa}
\usepackage[pdftex]{hyperref}
\usepackage{subfigure}
\usepackage{natbib}
\usepackage{helvet,soul}
\usepackage[font=small]{caption}

\contriblanguage{1}

\contribtype{1}

\thematicarea{7}

\title{Hard diffuse X-ray emission around the PSR\,J2032+4127:}
\subtitle{A pulsar wind nebula in the Cygnus\,OB2 association}

\titlerunning{A PWN in the Cygnus\,OB2 association}

\author{J.F. Albacete Colombo\inst{1}, J.J. Drake\inst{2}, A. Filócomo\inst{1} \& N.J. Wright\inst{3}}
\authorrunning{Albacete Colombo et al.}

\contact{jfalbacetecolombo@unrn.edu.ar}

\institute{
Dep. de Investigaci\'on en Ciencias Exactas e Ingenie\'ia, UNRN - Sede Atl\'antica, Viedma (8500), Argentina
\and
Smithsonian Astrophysical Observatory, 60 Garden St., Cambridge, MA 02138, USA.
\and 
Astrophysics Group, Keele University, Keele, ST5 5BG, UK.
}

\resumen{La región de Cygnus\,OB2, de $\sim 3.5-5\times 10 ^{6}$ años de edad, contiene una de las poblaciones estelares de alta masa más grandes de la Vía Láctea. Estas estrellas contribuyen a la emisión difusa observada a gran escala en rayos X blandos ($<2~\text{keV}$).  Detectamos además emisión difusa en rayos X duros ($>3~\text{keV}$) en direccion del pulsar PSR\,J2032+4127. Esta emisión elongada de forma toroidal abarca una estructura tipo \emph{jet} de tamaño $\approx 3'\times 2'$, y es coincidente con la posición de la fuente de rayos-$\gamma$ 4FGL\,J2032.3+4127. Confirmamos que la emisión difusa de rayos X duros es la nebulosa pulsante originada por el pulsar PSR\,J2032+4127, como consecuancia de una explosion de SN tipo \emph{core-collapse} en la región.}

\abstract{The Cygnus\,OB2 region, $\sim 3.5-5~\text{Myr}$ old, contains one of the most significant populations of massive stars of the Milky Way. Such stars substantially contribute to producing large scale soft ($<2~\text{keV}$) diffuse X-ray emission. We also detected hard ($>3~\text{keV}$) diffuse X-ray emission in the direction of the pulsar PSR\,J2032+4127. The torus-shaped emission spans a $\approx 3'\times 2'$ jet-like structure. It is spatially coincident with the Fermi $\gamma$-ray source 4FGL\,J2032.3+4127. We suggest that the hard diffuse X-ray emission is the pulsar wind nebula bearing the pulsar PSR\,J2032+4127, a consequence of a past core-collapse SN explosion in the region.}


\begin{document}
\maketitle
\section{Introduction}
\label{S_intro}

The search for supernova remnants (SNRs) in massive star-forming regions and young stellar associations provides valuable constraints for stellar evolutionary models towards the upper mass limit of stars. Of the variety of supernova (SN) explosion types, core-collapse ones leave a pulsar as debris of its evolutionary path. Pulsars accelerate particles such as electrons and positrons to ultra-relativistic energies that emit synchrotron radiation from the radio to soft $\gamma$-rays, and energize lower-energy photons by Inverse Compton (IC) scattering up to TeV energies. However, not all neutron stars manifest themselves through the creation of a pulsar wind nebula (PWN), so their detection becomes essential to understand high energy processes on core-collapse SNRs.

The TeV J2032+4130 source was the first unidentified $\gamma$-ray detection by the HEGRA experiment. Interestingly, its position overlaps the edge of the 95\% confidence ellipse of the 3EG\,J2033+4118 EGRET source. However, it is still not clear if they are associated or not.
The massive Cygnus\,OB2 association lies at a distance of about $1.45-1.7~\text{kpc}$ \citep{Hanson2003, berl19}, and the TeV J2032+4130 $\gamma$-ray source in alignment with the suspected background pulsar PSR\,J2032+4127. Several studies discuss that the pulsar probably belongs to Cygnus\,OB2, and would be considered the counterpart of the TeV J2032+4130 source \citep{Butt2006, Horns2007}. More recently, \cite{Camilo2009} suggests that PSR\,J2032+4127 is probably one of the least energetic TeV pulsars powering faint or unconfirmed PWN. \cite{Mukherjee2007} noted its location, projected close to the core of several massive stars, and suggests a distance of 1.6~kpc, rather than 3.6~kpc estimated from the radio pulsar dispersion \citep{Cordes2002}. The existence of a neutron star in Cygnus\,OB2 would be consistent with an age of $1-7~\text{Myr}$ for the association \citep{Wright2010}. Notably, \cite{Wright2015a} found a steepening of the IMF slope at higher masses that they interpreted as due to an expected previous generation of "lost" massive stars, opening the possibility for some of these massive stars to have exploded as SNe in the past.

Finally, our recent analysis of diffuse X-ray emission in the Cygnus\,OB2 region \citep{Albacete-Colombo2018}, confirms now the first true detection of hard X-ray diffuse emission around  PSR\,J2032+4127, which also has positional agreement with the 4FGL\,J2032.2+4127 Fermi $\gamma$-ray source. 

\begin{figure*}[!t]
\centering
\includegraphics[width=0.99\textwidth]{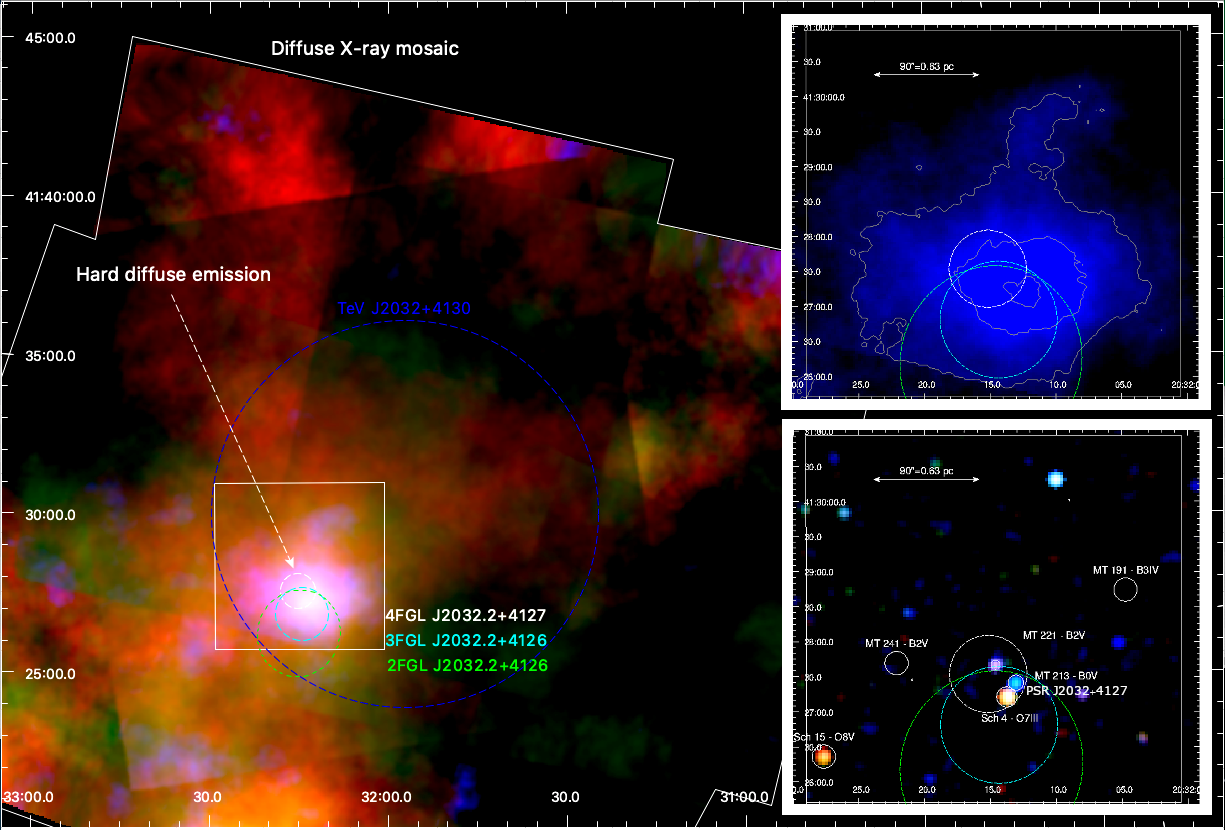}
\caption{Left panel: Chandra ACIS-I diffuse X-ray emission images around the TeV J2032+4130 pulsar. The mosaic was constructed by combining point source-removed observations and using an adaptive top-hat smoothing at $\text{S/N}>16$ \citep{Albacete-Colombo2018}. The X-ray images of the different bands are shown in color, with soft ($0.5-1.2~\text{keV}$), medium ($1.2-2.5~\text{keV}$), and hard ($2.5-7.0~\text{keV}$) emission indicated in red, green, and blue, respectively. The dashed line circles refer to the TeV and 2FGL (green), 3FGL (cyan), and 4FGL (white) Fermi source position ellipses, respectively. Hard diffuse X-ray emission appears nearly centered at the 4FGL source. Upper-right panel: Zoomed hard diffuse X-ray emission map with contour levels at factors 1.28, 1.55, and 1.82 over the hard X-ray flux background level $\approx 1.6 \times 10^{-9}~{\rm ph\,s^{-1}\,cm^{-2}}$. Bottom-right panel: RGB color-coded point source ACIS-I observation. The hard X-ray point sources refer to both the PSR\,J2032+4127 and MT213 (B0V) sources. The pulsar is not dynamically related to the MT91\,\#213 Be star, and is in coincident projection on the sky \citep{Camilo2009}, so the observed hard X-ray emission is likely consistent with the pulsar emission rather than the X-ray emission from shocks in weak-winds of the B0\,V massive star.}
\label{diffuse+fermi}
\end{figure*}

\section{X/$\gamma$-ray Observations}
\label{S_obs}
In the context of the Chandra Cygnus\,OB2 Large Program (PI: J.J. Drake), we use five Chandra pointings (ObsId 4501, 10944, 10945, 10951, 10962) that cover the position of the PSR\,J2032+4127 source. These observations span a total exposure time of $165.7~\text{ks}$, 3.3 times deeper than  previous diffuse X-ray emission studies \citep{Mukherjee2007}. Our improved X-ray diffuse analysis made use of the sophisticated  Acis-Extract software \citep{Broos2012} that removes the PSFs of all point-sources in the observations and constructs background-corrected diffuse X-ray maps \citep{Albacete-Colombo2018}. So, we have a unique observational advantage to confirm the existence of truly diffuse features missed in previous works that smooth observations that include point sources. Figure~\ref{diffuse+fermi} shows a mosaic of diffuse X-ray emission that includes TeV\,J2032+4130 in the  field-of-view.

The HEGRA and MAGIC $\gamma$-ray observatories detect significant emission (from MeV to TeV energies) at 0.2' of the core of the Cygnus\,OB2 stellar association (see Table~1). However, it has a poor determination of the position with respect to the last 4FGL Fermi catalog.

\begin{table}[!t]
\begin{center}
\caption{X/$\gamma$ ray sources associated to PSR\,J2032+4127.
Flux values are informed in units of 
$\Phi_{11}\equiv 10^{-11}~{\rm erg\,cm^{-2}}$ ${\rm s^{-1}}$. 
The HEGRA and MAGIC fluxes are in the energy range $\sim 1-10^4~\text{TeV}$ and $\sim 30~\text{GeV}-100~\text{TeV}$, respectively. The Fermi energy flux is in the range $100~\text{MeV} - 100~\text{GeV}$}.
\begin{tabular}{lcccc}
\hline\hline\noalign{\smallskip}
\!\!Mission & \!\!\!\!Name & \!\!\!\! Flux  & $\Gamma$\!\!\!\!& Dist.\\
    Name    & \!\!\!\!catalogue & \!\!\!\!$[\Phi_{11}]$ & \!\!\!\!& [ ' ] \\
\hline\noalign{\smallskip}
HEGRA  &  TeV\,J2032+4130 & 6.9 & 1.9 & 4.4\\
MAGIC  &  TeV\,J2032+4130 & 4.5 & 2.0& 1.2\\
4FGL  & J2032.2+4127 & 14.2$\pm$0.04 & 2.2 & 0.2\\
Chandra& PWN                & 0.02$\pm$0.05& 2.1 & --- \\
\hline
\end{tabular}
\end{center}
\label{T_param}
\end{table}

\section{Discussion}
\label{S_dis}

Two main observational issues that strongly suggest our detection of hard diffuse X-ray emission is  the PWN of the pulsar PSR\,J2032+4127, belonging the Cygnus\,OB2 association.

A preliminary study by \cite{Murakami2011} made use of Suzaku and Chandra data to shows that extended diffuse emission around PSR\,J2032+4127 has a non-thermal X-ray spectrum $\Gamma=2.1\pm 0.3$ (see Table~1), similar to typical $\Gamma$ indices of known PWN \citep{Kargaltsev2008}. However, the poor spatial resolution of Suzaku, the lack of spatial disentangling between the pulsar emission and its PWN, and the short exposure time ($49~\text{ksec}$) of the Chandra data, hamper the results.

In Figure~\ref{diffuse+fermi}, we show extended soft and medium band emission that spatially superposes the hard diffuse emission around the pulsar. While the softer emission arises from the cumulative interaction of energetic termination shocks against the surrounding interstellar medium \citep{Albacete-Colombo2018}, hard X-ray emission shows a jet-like morphology around the pulsar PSR\,J2032+4127 that agree with the detection of the 4FGL\,J2032.2+4127 Fermi and the TeV\,J2032+4130 $\gamma$-ray sources. Otherwise, the observed hard diffuse X-ray emission also matches the positions of the soft X-ray sources Schulte\,\#4 (O7I+B0V), and MT91\#213 (B0V) stars, which are unlikely to produce the observed diffuse hard emission via stellar wind-wind interactions. The elongated morphology along the pulsar axis-like extension is 1.3~pc, with a toroidal-like flow of 0.9~pc, which is typical in pulsars wind nebulae. The extension of both features agrees with the characteristic size of several known PWN that range between $10^{17}$ and $10^{19}~\text{cm}$ \citep{Cheng2004}.

\begin{figure}[ht]
\centering
\includegraphics[width=0.5\textwidth]{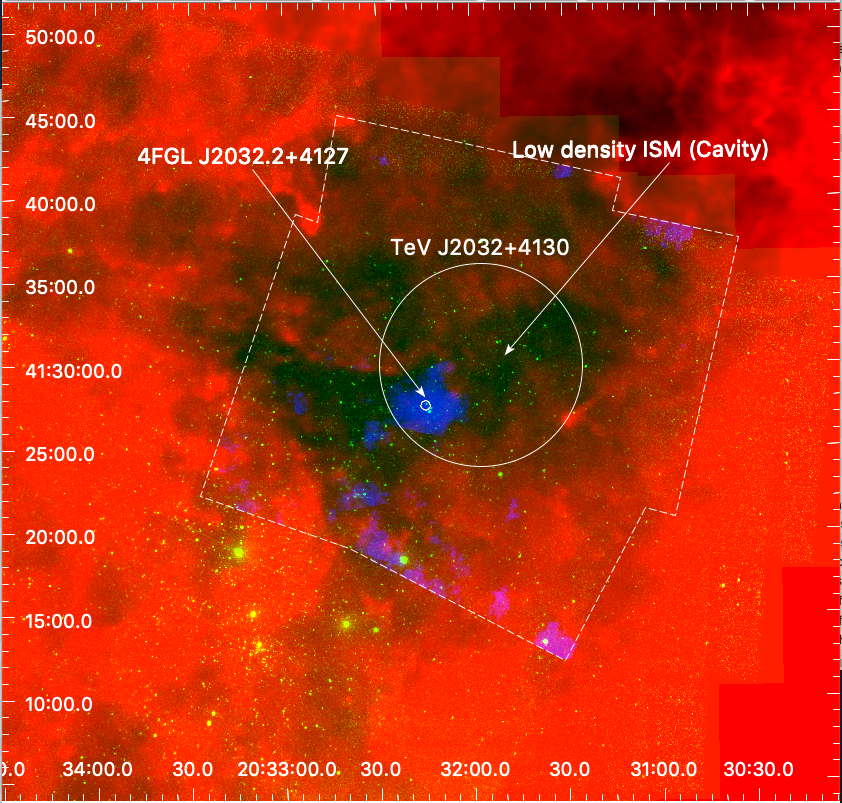}
\caption{RGB color image of the region in the vicinity of PSR J2032+4127. In the compose image, the Herschel $500~\mu\text{m}$ ($T\approx 10~\text{K}$) cold gas emission is indicated in red \citep{Schneider2016}, the Chandra point-source X-ray mosaic ($0.5-7.0~\text{keV}$) is in green, and the diffuse hard X-ray band ($2.5-7.0~\text{keV}$) in blue. Large and small position error ellipses refer to the TeV\,J2032+4130 and the Fermi 4FGL\,J2032.2+4127 sources, respectively. Notice how the low cold-ISM density correlates with the hard diffuse X-ray emission and the PSR\,J2032+4127 source position (see discussion). The cavity has an extension of $\approx 26'$ ($\sim 11~\text{pc}$).}
\label{hole}
\end{figure}

A hitherto unnoticed observational issue is the suggestive low-density (cavity) of cold gas and dust around the PSR\,J2032+4127, and its PWN (see Fig.~\ref{hole}). Indeed, the chance positional coincidence of the pulsar with this cavity is unlikely. The only known process to disperse the ISM on parsec scales is through a shock wave originating in a SN explosion. A core-collapse SN explosion in the region would be the most logical scenario compatible with an ISM cavity and a pulsar and PWN inside. Indeed, the lack of X-ray emission from a hypothetical SNR in Cygnus\,OB2 does not weaken our interpretation, as the radiative phase in SNRs ends after 20\,000 to 35\,000~yr from the explosion \citep{Slane2015}. Otherwise, the X-ray luminosity decreases with the pulsar age because of radiative cooling in the earlier stage ($0.1-1~\text{Myr}$), while the $\gamma$-ray luminosity remains constant. So, all these constraints are in agreement with a probable SN explosion in the past of Cygnus\,OB2, especially if we accept the characteristic age of the pulsar PSR\,J2032+4127 is $\approx 0.11~\text{Myr}$ \citep{Camilo2009}.

\section{Conclusion}
\label{S_conc}

Previous theoretical results and the observational constraints of this work are consistent with a $\sim 10^5~\text{yr}$ old energetic core-collapse SN explosion in Cygnus\,OB2. The scenario could be the result of a shock blast moving out and dispersing the surrounding ISM, leaving a pulsar at the center of an apparent cavity. In such a case, the pulsar PSR\,J2032+4127 would become the most evolved stellar member of the Cygnus\,OB2 region, and the detected diffuse hard X-ray emission its PWN.

\begin{acknowledgement}
We thank all LOC and SOC members for the successful meeting held in Viedma, September 2019, Rio Negro, Argentina.  JFAC is a staff researcher of the CONICET and acknowledges support from UNRN (PI 40-C-691), CONICET (PIP-0102), and ANPCyT (PICT-2017-2865).
\end{acknowledgement}


\bibliographystyle{baaa}
\small
\bibliography{facundo.bib}
 
\end{document}